\documentclass[preprint,onecolumn,showpacs,preprintnumbers,amsmath,amssymb]{revtex4}
\usepackage{epsf}
\usepackage{dcolumn}
\usepackage{bm}
\everymath{\displaystyle}
\begin{document}

\title{Connection between beam polarization and systematical errors in storage ring electric-dipole-moment experiments}
\author{Alexander J. Silenko}
\affiliation{Research Institute for Nuclear Problems, Belarusian State University, Minsk 220030, Belarus\\
Bogoliubov Laboratory of Theoretical Physics, Joint Institute for Nuclear Research,
Dubna 141980, Russia}

\date{\today}

\begin {abstract}
Analysis of spin dynamics in storage ring electric-dipole-moment (EDM) experiments ascertains that the use of initial vertical beam polarization allows to cancel spin-dependent systematical errors imitating the EDM effect. While the use of this polarization meets certain difficulties, it should be considered as an alternative or supplementary possibility of fulfilling the EDM experiment.
\end{abstract}

\maketitle

\section{Introduction}

Experiments to search for the EDMs are essential for the modern physics. Storage ring EDM experiments are very sensitive and can be fulfilled by two main methods. The frozen spin method \cite{FJM,AIP} almost cancels the \emph{g}-2 spin precession thanks to an appropriate radial electric field. The resonance method \cite{OMS,Lehrach:2012eg} allows a modulation of the electric field in the particle rest frame. For this purpose, one can modulate a beam velocity \cite{OMS}, an electric field \cite{Lehrach:2012eg}, or a magnetic one \cite{Lehrach:2012eg} in the lab frame. Both of the methods create a nonzero average radial electric field acting on the EDM in the particle rest frame. Since an oscillating field of a flipper covers a little part of a beam trajectory, the frozen spin method ensures a higher precision.  The storage ring EDM experiments based on the frozen spin method are planned for muons \cite{FJM,muEDM,compactmuEDM}, deuterons \cite{AIP,dEDM}, and protons \cite{AIP}.

In all planned frozen-spin storage ring EDM experiments, the initial beam polarization is scheduled to be longitudinal. In the present work, we describe the spin dynamics in detail and consider an alternative possibility to use the initial longitudinal polarization of the beam. We prove that the latter polarization cancel spin-dependent systematical errors imitating the EDM effect while its use can bring other difficulties.

\section{Spin dynamics in storage ring electric-dipole-moment experiments}

Spin dynamics in the planned frozen-spin EDM experiments with muon, deuteron, and proton beams has both similarities and differences. While the spin behavior caused by the EDM is rather specific, a similar behavior can be stimulated by tensor polarizabilities (for the deuteron) and main systematical errors originated from field distortions and misalignments. For the initial longitudinal polarization of the beam, the interactions caused by the EDM, tensor polarizabilities, and main systematical errors result in the effect of the buildup of the vertical polarization (BVP). These interactions can nevertheless be distinguished because they have different symmetries \cite{dEDMtensor,dEDMtnsr}.

Their distinguishing is a difficult problem of searching a regular BVP against the background.
The \emph{g}-2 spin precession in the horizontal plane takes place in any storage ring EDM experiments. Even if the frozen spin method is applied and the \emph{g}-2 spin precession is vanished on the average, a residual precession exists. In this case, the direction of the spin rotation depends on a particle momentum and can differ for different particles. Therefore, it can be preferable to leave a slow average \emph{g}-2 spin precession of the beam  \cite{AIP,dEDM}. In the deuteron EDM experiment, one would have up to three runs in every clockwise (CW)/counter-clockwise (CCW) injection, with variable control of the \emph{g}-2 precession rate, so that at the end of the three $10^3$ s running periods the spin will make about 5, 2 and 1/2 horizontal turns, respectively \cite{dEDM}. As a result, the change of the phase, $\omega_0t$, is not small.

In the general case, the spin dynamics is described by Eq. (22) in Ref. \cite{dEDMtensor}:
\begin{equation}
\begin{array}{c}
P_{\rho}(t)=\frac{{\omega_0\cal
C}}{{\omega'}^2}\left[1-\cos{(\omega't)}\right]\cos{\theta}
+\!\left[1\!-\!\frac{2\omega_0^2}{{\omega'}^2}\sin^2{\frac{\omega't}{2}}\right]\sin{\theta}
\cos{\psi}\\-\frac{\omega_0}{\omega'}\sin{(\omega't)}\sin{\theta}\sin{\psi},\\
P_{\phi}(t)=\sin{(\omega't)}\left(\frac{\omega_0}{\omega'}
\sin{\theta}\cos{\psi}-\frac{{\cal
C}}{\omega'}\cos{\theta}\right)+\cos{(\omega't)}
\sin{\theta}\sin{\psi},\\
P_{z}(t)=\left[1-\frac{2{\cal
C}^2}{{\omega'}^2}\sin^2{\frac{\omega't}{2}}\right]\cos{\theta}
+\frac{{\omega_0\cal
C}}{{\omega'}^2}\left[1-\cos{(\omega't)}\right]\sin{\theta}
\cos{\psi}\\+\frac{{\cal
C}}{\omega'}\sin{(\omega't)}\sin{\theta}\sin{\psi},
\end{array}
\label{propedm}
\end{equation}
where
\begin{equation}
\bm\omega'=\omega_0\bm e_z+{\cal
C}\bm e_\rho, ~~~ {\cal
C}=-\frac{e\eta}{2m}\cdot\frac{1+a}{1-a\beta^2\gamma^2}\beta_\phi
B_z,
\label{defnedm}
\end{equation}
$\bm\omega'$ and $\bm\omega_0=\omega_0\bm e_z$ are the angular velocities of spin precession of particle with and without the EDM, respectively, $\theta$ and $\psi$ are the angles of the spherical coordinate system defining the direction of the initial polarization of the beam, $\bm\beta\equiv\bm v/c$, $\bm P$ is the polarization vector, $\eta=2dm/(es)$, and $d$ is the EDM. Equations (\ref{propedm}),(\ref{defnedm}) show that each of the horizontal spin components ensures the same level of information about the EDM.

When the tensor polarizabilities are not taken into account (for the deuteron), the
spin rotates about the direction
$$\bm e_z'=\frac{{\cal C}}{\omega'}\bm e_\rho+\frac{\omega_0}{\omega'}\bm e_z$$
with the angular frequency $\omega'=\sqrt{\omega_0^2+{\cal C}^2}$. As a rule, ${\cal C}$ is much smaller than $\omega_0$.

The effect of tensor polarizabilities on the spin dynamics of the deuteron has been calculated in Refs. \cite{dEDMtensor,dEDMtnsr}.

For the three initial polarizations, longitudinal, radial, and vertical, the EDM effect has different manifestations and the same order of magnitude. The maximum BVP is $P_z^{(max)}={\cal C}/\omega'\approx{\cal C}/\omega_0$ and $P_z^{(max)}=2\omega_0{\cal
C}/{\omega'}^2\approx2{\cal C}/\omega_0$ for the initial longitudinal and radial polarization, respectively. The amplitudes of the vertical oscillations of the spin are almost equal in the two cases: $(P_z^{(max)}-P_z^{(min)})/2
\approx{\cal C}/\omega_0$. When the initial beam polarization is vertical ($\theta=0$), the spin rotates in the horizontal plane with the small amplitude, ${\cal C}/\omega_0$, proportional to the EDM. The radial and longitudinal spin components oscillate between $2\omega_0{\cal
C}/{\omega'}^2$ and 0 and between ${\cal C}/\omega'$ and $-{\cal C}/\omega'$, respectively. When the initial polarization is radial or vertical, the minor spin components (vertical and radial, respectively) caused by the EDM are nonzero on average.

In all the considered cases, the phase is rather informative. When the initial polarization is longitudinal and $\omega_0t\ll1$, the BVP is equal to
\begin{equation}
P_z={\cal
C}t.
\label{eqthree}
\end{equation}
When the initial polarization is radial, the corresponding BVP is very small and equal to $\omega_0{\cal
C}t^2/2$. When the initial polarization is vertical, the longitudinal final polarization is equal to ${\cal C}t$ and the radial one is very small and equal to $\omega_0{\cal
C}t^2/2$.

\section{Spin dynamics caused by main systematical errors. Extracting the electric-dipole-moment signal}

It is important that the spin interactions depending on the EDM, the tensor electric and magnetic polarizabilities (for the deuteron), and main (first-order) systematical errors caused by field distortions and misalignments have very different symmetries. The differences of symmetries take place for any direction of the initial polarization or the beam. The minor spin component caused by the EDM changes the sign after reversing the initial polarization and conserves the sign after reversing the beam direction. There are two kinds of systematical errors which affect the spin and can imitate the EDM signal. Main systematical errors are of the first order in the field strengths (first-order systematical errors). These errors change the sign of the spin deflection after reversing either the initial polarization or the beam direction. As a result, they can be canceled when two beams moving in CW and CCW directions \cite{dEDM}. Other systematical errors are of the second order in the field strengths (second-order systematical errors, also called geometrical phase and Berry's phase effects). They appear due to a non-commutativity of spin rotations about different axes \cite{dEDM}. Such systematical errors are much less than the first-order one. They are nevertheless very undesirable because only some of them seem as $T$-even and can be canceled with CW and CCW beams \cite{note1,note2}. Other second-order systematical errors cannot be canceled with CW/CCW beams and seem as $T$-odd. One can use the property that all the second-order systematical errors differ for the longitudinal and radial initial polarizations. Potential for an elimination of ``$T$-odd'' second-order systematical errors has been considered in Ref. \cite{dEDM}. However, a development of new methods of their elimination remains important.

Systematical errors may cause the spin rotation about the both radial ($\bm e_\rho$) and longitudinal ($\bm e_\phi$) axes. The first-order systematical errors are conditioned by the vertical electric field and the radial magnetic one which are nonzero on average. These errors govern the spin rotation about the radial axis. In connection with the Maxwell equations, an average longitudinal magnetic field may appear only due to a nonzero total current inside the beam orbit. The correction caused by the Earth’s gravity has been calculated in Refs. \cite{PRD2,OFS}. This correction can also be canceled with CW/CCW beams.

The effect of the tensor electric and magnetic polarizabilities of the deuteron has been discussed in Refs. \cite{dEDMtensor,dEDMtnsr}. The tensor polarizabilities conserve the final sign of the minor spin component after reversing the initial polarization and the direction of the beam.
It is important that the EDM experiments in storage rings can be effectively used for high-precision measurements of the tensor polarizabilities of the deuteron and other nuclei.
The tensor electric polarizability can be determined with summing up the data for two opposite states of initial polarization. To find the tensor magnetic polarizability of the deuteron, the horizontal components of the polarization vector should be measured. The deuteron EDM can be determined with summing up the data for two opposite beam directions and taking into account the correction for the tensor electric and/or magnetic polarizabilities (see Refs. \cite{dEDMtensor,dEDMtnsr} and references therein).

It can be proved that the use of the vertical initial beam polarization cancels all second-order systematical errors. Any spin rotation is governed by external fields. An instantaneous angular velocity of the spin rotation, $\bm\omega'(t)$, is defined by the Thomas-Bargmann-Mishel-Telegdi equation added by the EDM-dependent terms \cite{Nelson:1959zz,NPRRPJSTAB,Spinunit}:
\begin{equation}
\begin{array}{c}
\bm\omega'=\bm\omega_{T-BMT}+\bm\omega_{EDM},\\
\bm\omega_{T-BMT}=-\frac{e}{m}\left[\left(\frac{g-2}{2}+\frac1\gamma\right)\bm B \right.\\ \left.-\frac{(g-2)\gamma}{2(\gamma+1)}\bm\beta(\bm\beta\cdot\bm B)-\left(\frac{g-2}{2}+\frac{1}{\gamma+1}\right)(\bm\beta\times\bm E)\right],\\
\bm\omega_{EDM}=-\frac{e\eta}{2m}\left[\bm E-\frac{\gamma}{\gamma+1}\bm\beta(\bm\beta\cdot\bm E)+\bm\beta\times\bm B\right].
\end{array}
\label{tbmtedm}\end{equation} 
The dependence of $\bm\omega'(t)$ from the electric and magnetic fields is linear.

The motion of the spin is given by
\begin{equation}
\Delta\bm s\equiv\bm s(t)-\bm s(0)=\int^t_0{\bm\omega'(t')\times\bm s(t')dt'}.
\label{spm}
\end{equation} We suppose $|\bm s|=1$.

When $|\Delta\bm s|\ll1$, we can assume $\bm s(t')\approx\bm s(0)$. In this case
\begin{equation}
\Delta\bm s=\int^t_0{\bm\omega'(t')dt'}\times\bm s(0)=\overline{\bm\omega'(t)}\times\bm s(0)\Delta t,
\label{aspm}
\end{equation}
where the overline denotes averaging. Just this situation takes place when the initial beam polarization is vertical. Equations (\ref{tbmtedm}),(\ref{aspm}) show that the spin deflection from the vertical, $\Delta\bm s$, is proportional to the average components of the electric and magnetic fields. The average components define the first-order systematical errors. Second-order systematical errors can appear due to a coupling of fluctuations of
the spin direction and those of the angular velocity of spin precession. Since the both fluctuations are proportional to the field strengths, the second-order systematical errors are bilinear in the external fields. The use of the vertical initial polarization allows to neglect fluctuations of
the spin direction and therefore cancels all second-order systematical errors.

The initial  vertical beam polarization allows a separation in space of the magnetic field of any bending magnet from the radial electric field. For the longitudinal and radial initial beam polarizations, such a separation results in the local \emph{g}-2 rotation which may also lead to the second-order systematical errors imitating the EDM \cite{dEDM}.

\section{Special features of muon electric-dipole-moment experiment}

The muon EDM experiment performed by the frozen spin method possesses some special features because of the short lifetime of the muons. In this case, the problem of the spin coherence is not so important than in the deuteron and proton experiments. The dynamics of the polarization vector of decaying muons is described by Eq. (\ref{propedm}). However, one is to consider just the polarization vector components,
$$P_i(t)=N_i(t)\Bigl/\sum^3_{i=1}N_i(t),$$ for each detector but not the number of decaying muons with the definite spin projection, $N_i(t)$. When one averages $P_i(t)$ (but not $N_i(t)$) over the beam circumference and the initial beam polarization is vertical, one cancels all second-order spin-dependent systematical errors.

The muon spin orientation is reconstructed from the distribution of the decay positrons. The simplest and most straightforward detection system only distinguishes upward and downward going positrons \cite{compactmuEDM}. Due to the magnetic field of the storage ring, the positron trajectories will be bent toward the inner side of the ring. Therefore, the detection of the inward-outward spin asymmetry may be more difficult. The relativistic motion of the muons complicates the detection of the forward-backward spin asymmetry. As a result, the measurement of the radial and longitudinal components of the polarization vector seems to be very difficult. However, one can restrict oneself to the determination of the time evolution of these components for any detector. If the time evolution is determined with a needed accuracy, one may disregard the position of the polarization vector at $t=0$. An additional use of the radially or longitudinally polarized beam bunches in every run seems to be appropriate.

\section{Discussion and summary}

If the vertical initial polarization is used, one can alternate the beam bunches with the upward and downward polarizations. An additional use of slow but nonzero \emph{g}-2 rotation can also be helpful. Equation (\ref{propedm}) shows that the \emph{g}-2 rotation changes the direction of the final horizontal polarization. This effect can be used while the \emph{g}-2 rotation decreases the EDM effect.

Monitoring of the beam does not allow to measure the \emph{g}-2 frequency and the spin coherence time when the initial beam polarization is vertical. Therefore, the use of the horizontal initial beam polarization is also necessary \cite{Semertzidis}. It is optimal to apply some horizontally polarized beam bunches for feedback on the machine parameters and some vertically polarized beam bunches for the EDM measurement. Any polarization can be applied in order to check for the rather seldom effect of spin flip 180 degrees.

When the initial beam polarization is longitudinal or radial, the slow but nonzero \emph{g}-2 rotation can be useful. In this case, the monitoring of the spin dynamics allows to measure the \emph{g}-2 frequency, check for the spin coherence, and separate spin rotations about the radial and longitudinal axes. These measurements cannot be fulfilled with the vertically polarized beam.

Unfortunately, a practical use of the vertical polarization in the proton and deuteron EDM experiments performed by the frozen spin method meets serious problems due to polarimeter systematic errors. They appear owing to changes of motion and inclination of the beam relatively the target during the storage time \cite{Semertzidis}. To apply this polarization in the above mentioned experiments, one need to solve the polarimetry problem.

The vertical initial polarization of the beam can be successfully applied in the resonance EDM experiments in storage rings. Indeed, the use of this polarization is planned in the proton and deuteron EDM experiments performed by the resonance method \cite{Lehrach:2012eg}. The problem of systematical errors in such experiments requires a separate consideration.

Thus, the use of the vertical initial polarization of the beam in storage ring electric-dipole-moment experiments allows to cancel the most dangerous systematical errors imitating the EDM effect.
While its application meets certain difficulties, it should be considered as an alternative or supplementary possibility of performing the EDM experiment.

\section*{Acknowledgements}

The author is indebted to F.\,J.\,M. Farley, J.\,P. Miller, W.\,M. Morse, F. Rathmann, Y.\,K. Semertzidis and other colleagues from the Storage Ring EDM and JEDI collaborations for valuable comments and discussions. The work was supported by the
Belarusian Republican Foundation for Fundamental Research
(Grant No. $\Phi$12D-002).


\end{document}